\documentstyle[epsf]{l-aa}
\begin{document}

\newcommand{\greq}{\begin{equation}\left\{ \begin{array}{l}}
\newcommand{\egreq}{\end{array}\right. \end{equation}}
\newcommand{\egreqn}[1]{\end{array}\right. \label{#1}\end{equation}}
\newcommand{\beqa}{\begin{eqnarray*}}
\newcommand{\eeqa}{\end{eqnarray*}}
\newcommand{\beqan}{\begin{eqnarray}}
\newcommand{\eeqan}[1]{\label{#1}\end{eqnarray}}
\newcommand{\beq}{\begin{equation}}
\newcommand{\eeq}{\end{equation}}
\newcommand{\eeqn}[1]{\label{#1}\end{equation}}
\newcommand{\beqm} {\begin{displaymath}}
\newcommand{\eeqm} {\end{displaymath}}
\newcommand{\diff}{{\rm d}}

\bibliographystyle{plain}

\thesaurus{
02.08.1; 
06.09.1, 
06.18.2; 
08.09.3, 
08.18.1. 
}

\title{Angular momentum transport by internal waves in the solar interior}


\author{Jean-Paul Zahn\inst{1}, Suzanne Talon\inst{1} and Jos\'e
Matias\inst{1,2}}

\offprints{Jean-Paul Zahn}

\institute{
D\'epartement d'Astrophysique Stellaire et Galactique, Observatoire de Paris, 
Section de Meudon, 92195 Meudon, France
\and Centro de Astrof\'{\i}sica, Universidade do Porto, Rua do Campo
Alegre 823, 415 Porto, Portugal \\
e-mail : zahn@obspm.fr, talon@obspm.fr, Jose.Matias@mail.telepac.pt}


\date{}

\maketitle
\markboth{J.-P. Zahn, S. Talon
and J. Matias: Angular momentum transport by internal waves}{}

\begin{abstract}
The internal gravity waves of low frequency which are emitted at the base of
the solar convection zone are able to extract angular momentum from the
radiative interior. We evaluate this transport with some simplifying
assumptions: we ignore the Coriolis force, approximate the spectrum of
turbulent convection by the Kolmogorov law, and couple this turbulence to
the  internal waves through their pressure fluctuations, following Press
(1981) and Garc\'{\i}a L\'opez \& Spruit (1991). The local frequency of an
internal wave varies with depth in a differentially rotating star,
and it can vanish at some location, thus leading to enhanced damping
(Goldreich \& Nicholson 1989). 
It is this dissipation mechanism only that we take into account in 
the exchange of momentum between waves and stellar rotation. The flux
of angular momentum is then an implicit function of depth, involving the local
rotation rate and an integral representing the cumulative effect of radiative
dissipation. We find that the efficiency of this transport process is rather
high: it operates on a timescale of 
$10^7$
years, and is probably
responsible for the flat rotation profile which has been detected through
helioseismology. 

\end{abstract}

\section{Introduction}
Although it is well known that waves do carry momentum, this
process has received little attention so far in stellar physics.The transport of
angular momentum by internal waves (also called gravity waves) has been
studied in the context of tidal braking involving massive binary stars (Zahn
1975a; Goldreich \& Nicholson 1989), but only recently has it been invoked as a
mechanism  which could shape the rotation profile in the Sun (Zahn 1990;
Schatzman 1993). In contrast, the importance of the momentum
transport by such waves has been recognized long ago in atmospheric
sciences (cf. Bretherton 1969a,b): this phenomenon is responsible in
particular for the so-called clear air turbulence.

The purpose of the present paper is to assess the efficiency of
angular momentum transport in a solar-type star, where such waves are generated by
the turbulent motions of the convection zone. In this first approach we shall
make several simplifying assumptions, fully aware that the outcome must be
considered as a crude approximation. The results may be easily transposed to
massive stars, where these waves are emitted by the convective core.

The reason for examining the role of such waves in the Sun
is that the other mechanisms which have been analyzed so far seem unable to
achieve the flat rotation profile revealed by helioseismology (cf. Brown et
al. 1989). Both rotation-induced turbulent diffusion (Endal \& Sofia 1978;
Pinsonneault et al. 1989) and wind-driven meridian circulation (Zahn 1992)
fail to extract sufficient angular momentum from the radiative interior (Chaboyer
et al. 1995; Matias \& Zahn 1996). Magnetic torquing looks at first sight more
promising, but the field lines anchored in the differentially rotating convection
zone would probably enforce non uniform rotation below (cf. Charbonneau \&
MacGregor 1993), and this is not observed.

We begin by recalling the main properties of the internal waves (\S2), and
calculate the flux of angular momentum carried by a monochromatic wave (\S3). 
We then deduce the energy spectrum of those waves from their coupling with 
the turbulent motions at the base of the convection zone (\S4) and
integrate the angular momentum flux over the whole spectrum (\S5).
We finally derive an estimate for the efficiency of the angular momentum 
transport by such waves in the Sun (\S6) and end by some concluding remarks.

\section{Properties of internal waves}
The properties of internal waves propagating in stellar interiors have been
described by Press (1981) and by Gold\-reich \& Nicholson (1989). Let us recall
the main results, and adapt them to a differentially rotating star. We use the
spherical coordinates ($r$, $\theta$,  $\phi$) suited for this problem; $\vec
e_z$ is the unit vector on the rotation axis, and $\vec e_\phi$ that in the
azimuthal direction. We further assume that the angular  velocity $\Omega$
depends only on depth, because differential rotation in latitude is severely
limited through hydrodynamical instabilities (Zahn 1975b, 1992). 
The velocity field with respect to an inertial frame is then
\beq
\vec U (r,\theta, \phi, t)  = \Omega (r) \vec e_z  \times \vec r + 
\vec u (r,\theta, \phi, t),
\eeqn{totv}
with $\vec u$ being the velocity associated with the wave. 
The equation of motion reads 
\beqan
{\diff\vec u  \over \diff t}
&+& \Bigl\{ 2 \Omega \vec e_z \times \vec u + \vec e_\phi\, 
r \sin \theta \,\vec u
\cdot \vec \nabla \Omega \Bigr\} \nonumber \\
&=& - {1 \over \rho} \vec \nabla P' + {\rho ' \over \rho} \vec g  , 
\eeqan{motion}
with 
\beq
{\diff \over \diff t} = \left({\partial \over \partial t} + 
\Omega {\partial \over \partial \phi} \right) ,
\eeqn{doppler}
and the usual notations for the gravity $\vec g$, and for the Eulerian 
perturbations of pressure ($P'$) and density ($\rho'$).
We have simplified this equation by neglecting the fluctuations of the
gravitational potential: the Cowling approximation is amply justified here,
considering the high radial order of the waves. Viscosity also has been
ignored. We add the continuity equation  
\beq
{\diff \rho' \over \diff t}
 + \vec \nabla \cdot \rho \vec u = 0,
\eeqn{cont}
and the energy equation, which in the adiabatic limit reduces to
\beq
{\diff \over \diff t}
\left({\rho' \over \rho} - {1 \over \Gamma_1}{P' \over P}\right) 
+ \left[{\diff \ln \rho \over \diff r}- {1 \over \Gamma_1}{\diff \ln P
\over\diff r}\right] u_r = 0 ,
\eeqn{adiab}
$\Gamma_1$ being the adiabatic exponent.

We now proceed with a further simplification, which admittedly is much less
justified. Neglecting the terms in curly brackets in the equation of motion 
(\ref{motion}), we treat the waves as if they were pure gravity waves which 
are not modified by the  Coriolis acceleration, but just feel the 
entrainment by the differential rotation.

Then the differential system above is very similar to that governing the
internal waves in a non-rotating star; the only difference lies in the 
derivative with respect to time, which is replaced here by (\ref{doppler}).
The solutions are separable in spherical functions and time, as shown for
the vertical component of the velocity:
\beq
u_r(r, \theta, \phi, t) = u_v(r) P_\ell^m(\cos \theta) 
\exp i [\sigma t - m (\phi - \Omega t)], 
\eeqn{vvert}
where 
\beq
\sigma(r) = \sigma_0 - m \Omega(r) ,
\eeq
with $\sigma_0$ being the frequency in the inertial frame. 
In the low frequency range of the internal waves which are considered here, 
the function $\Psi (r) = \rho^{1\over2} r^2 u_v$ obeys the second order
equation (cf. Press 1981):  
\beq
{\diff^2 \Psi \over \diff r^2} + \Bigl({N^2\over\sigma^2} -1\Bigr)
{\ell(\ell+1)\over r^2} \Psi = 0 \,.
\eeqn{secorder} 
The buoyancy (or Brunt-V\"ais\"al\"a) frequency $N$ is given by 
\beq
N^2 = N_t^2 + N_\mu^2 = 
{g \delta \over H_P} (\nabla_{\rm ad} - \nabla) + 
{g \varphi \over H_P}  \nabla_\mu 
\eeq
where we have used the classical notations of stellar structure theory: 
$H_P$ is the pressure scale height $P / \rho g$, 
$\nabla = \partial \ln T / \partial \ln P$ the
logarithmic temperature gradient, 
$\delta = - (\partial \ln \rho /\partial \ln
T)_{P,\mu}$ and $\varphi = (\partial \ln \rho /\partial \ln \mu)_{P,T}$. 
We allow for a molecular weight gradient 
$\nabla_\mu = \diff \ln \mu / \diff \ln P$, whose
contribution may be important in some cases. Let us also recall that, due to
convective penetration, $N$ starts at a finite value $N_c$ at the top of the
radiation zone (Zahn 1991).

We introduce the vertical wavenumber $k_v$:
\beq
k_v^2 =  \left({N^2\over\sigma^2} -1\right) \, {\ell(\ell+1)\over r^2}
\eeqn{wavenb}
and observe that $r k_v \gg 1$ for $\sigma \ll N$. Therefore  the
differential equation (\ref{secorder}) may be solved by the WKB method, 
which yields the following result:  
\beqan
u_r &=& {\rm C} \; r^{-{3\over2}} \rho^{-{1 \over 2}}
\left({N^2\over\sigma^2} -1\right)^{-{1 \over 4}}
 P_\ell^m(\cos \theta)  \nonumber \\
&\times&\cos \left(\sigma t - m (\phi - \Omega t) - 
\int_r^{r_c} k_v \diff r \right),
\eeqan{wkbr}
with $r_c$ designating the base of the convection zone and
the constant C fixing the amplitude (cf. Press 1981). It describes a
``monochromatic'' wave of spherical order $\ell$ and local frequency 
$\sigma$, which propagates\footnote{The most general solution would include a
stationary wave.} 
with the phase velocity ($-\sigma/k_v$, $0$, $r
\sin \theta \, \sigma/m$), $m$ being the azimuthal order; 
its vertical group velocity is given by 
\beq
V_g = - {\diff \sigma \over \diff k_v} = {\sigma \over k_v} \;
{N^2 - \sigma^2 \over N^2}. 
 \eeqn{vgroup}

Making use of the continuity equation (\ref{cont}), we obtain similar 
expressions for the horizontal components of the velocity. In particular 
\beqan 
u_\phi &=& {\rm C}
\;  m \, { r k_v\over \ell(\ell+1)} \; r^{-{3\over2}} \rho^{-{1\over 2}} 
\left({N^2\over\sigma^2} -1\right)^{-{1 \over 4}}
 {P_\ell^m (\cos \theta) \over \sin \theta}  \nonumber \\
&\times& \cos \left(\sigma t - m (\phi - \Omega t) - \int_r^{r_c} k_v 
\diff r \right) \nonumber \\ &=&  m \, { r k_v\over \ell(\ell+1)}\; {u_r
\over \sin \theta}\; .  
\eeqan{wkbphi}

\section{Fluxes associated with a monochromatic wave}
The horizontal average of the kinetic energy density is easily deduced from 
the WKB solution above:
\beqan
{1\over2}\, \rho <u^2> &\equiv&  {1 \over 4 \pi} \int\!\!\!\int  {1\over2} 
\; \rho (u_r^2 + u_\theta^2 + u_\phi^2) \, \sin \theta \; \diff \theta \,
\diff \phi  \nonumber  \\ &=& {1\over2}\; {N^2 \over \sigma^2} 
{1 \over 4 \pi} \int\!\!\!\int \rho \; u_r^2 \, \sin \theta \; \diff \theta 
\, \diff \phi  \\ &\equiv&  {1\over2}\, {N^2 \over \sigma^2}\,
\rho\, <u_r^2>. \nonumber
\eeqan{endensity}
Multiplying by the group velocity (\ref{vgroup}), we get the average flux of kinetic
energy transported by a traveling wave: 
\beqan
{\cal F}_K &=& {1\over2} \, \rho <u^2> \,{\sigma \over k_v} 
{N^2 - \sigma^2 \over N^2} \nonumber \\
 &=&  {\rm sign}(k_v) \, {1\over2} \, \rho <u^2> \, {\sigma^2 \over N^2} 
\, {(N^2 - \sigma^2)^{1 \over 2} \over k_h} ,
\eeqan{fluxk}
with $k_h = \sqrt{\ell(\ell +1)} / r$ being the horizontal wavenumber. 

Likewise we evaluate the mean flux of angular momentum carried by the wave:
\beqan
{\cal F}_J &=& {1 \over 4 \pi} \int\!\!\!\int   \rho\, r \sin \theta\, 
u_\phi u_r \,  \sin \theta \; \diff \theta \, \diff \phi 
\nonumber  \\
&=&  m \,\rho r \,{r k_v \over \ell(\ell+1)}{1 \over 4 \pi} \int\!\!\!\int 
u_r^2 \, \sin \theta \; \diff \theta \, \diff \phi \label{fluxjex}  \\
&=& 2 \,{m \over \sigma} \; {\cal F}_K ,
\eeqan{fluxj}
where we have used (\ref{wavenb}), (\ref{wkbphi}) and (\ref{endensity}). 
Inserting (\ref{wkbr}) into (\ref{fluxjex}), we verify that the angular
momentum is conserved in this adiabatic limit\footnote{This condition
is not fulfilled in Schatzman (1993).}:
\beq
4 \pi r^2 {\cal F}_J(r) \equiv {\cal L}_J = \hbox{cst}.
\eeq
 In contrast, the kinetic
energy of the wave varies with depth, since the local frequency $\sigma$ 
depends on the rotation rate $\Omega(r)$, but the conservation of energy  
is ensured in the inertial frame, as was explained by Bretherton (1969a)
in the plane-parallel case.

When radiative damping is taken into account in the quasi-adiabatic limit, 
the wave amplitude is multiplied by an attenuation factor $\exp
(-\tau/2)$,      where $\tau$, which is similar to an optical depth, is
the integral  
\beq
\tau(r) = [\ell(\ell+1)]^{3\over2} \int_r^{r_c} K  \; {N N_t^2 \over
\sigma^4}  \left({N^2 \over N^2 - \sigma^2}\right)^{1 \over 2} {\diff r
\over r^3}, 
\eeqn{integral}
with $K$ being the thermal diffusivity (see Appendix B).
Then the angular momentum luminosity is no longer conserved, but decreases 
as \beq
{\cal L}_J(r) = {\cal L}_J(r_c)\,  \exp[ -\tau(r)],
\eeqn{dampj}
which means that some angular momentum will be extracted from the radiative
interior.
 
 \section{Spectral distribution}
From now on, we focus only on those waves which propagate towards the
radiative interior and thus carry energy upwards
($k_v >0$).
As we have seen above in (\ref{fluxk}), the average kinetic energy flux 
transported by such a wave, at the top of the radiation zone  
($r = r_c$), is given by   
\beq  
{\cal F}_K(r_c) = {1\over2} \, \rho \,{(N_c^2-\omega^2)^{1 \over 2} \over 
k_h} \, {\omega^2 \over N_c^2}\, {\overline u}^2,   
\eeqn{kfluxcbase} 
where ${\overline u}(\omega, k_h)$ is the r.m.s. of the wave velocity over 
the horizontal surface, $N_c$ the buoyancy frequency, and $\omega$ the 
frequency
of the wave in the frame rotating with the angular velocity ${\Omega_c}$ of
the convection  zone (Press 1981; Garc\'{\i}a L\'opez \& Spruit 1991).

We now turn to the evaluation of the flux carried by the whole spectrum of 
 waves generated by the convective motions. In all likelihood the excitation
of these internal waves takes place very close to the base of the
convection zone. There are two reasons for that. First the gravity waves do
not penetrate far into the convection zone, since they are evanescent there
with the lapse rate $k_h$. And second the interface is rather sharp, with a
jump in the buoyancy frequency and plumes penetrating into the radiative
interior, which favors the generation of waves as one observes both in the
laboratory (Townsend 1958) and in numerical simulations (Hurlburt et al.
1986, 1994). 

To obtain a crude estimate of the wave energy, we follow closely 
Garc\'{\i}a L\'opez \& Spruit (1991). We match the pressure fluctuation  
in the wave with that of the turbulent convection, thus  
\beq 
{\overline u}^2(\omega) = v^2(\omega), 
\eeq  
where $v(\omega)$ designates the r.m.s. convective velocity at the 
frequency $\omega$. We further take into account that convective eddies of
wavenumber $k(\omega)$ generate also waves at lower (horizontal) wavenumber
$k_h$. Summing the eddies which participate in the stochastic excitation 
of a wave of wavenumber $k_h$, we get 
\beq 
{\overline u}^2(\omega, k_h) = \left[{k_h \over k(\omega)}\right]^2 \!\!
v^2(\omega) \equiv 2 \!\int_0^{k_h}\!\! v^2(\omega) \!\left[{k_h \over
k(\omega)}\right]^2 \!{\diff k_h \over k_h} .  
\eeqn{distrik} 
 
Next we assume that the kinetic energy spectrum of the convective motions 
is represented to first approximation by the Kolmogorov law:
\beq 
v^2(\omega) = \int_\omega^\infty v_c^2 \left[\omega \over
\omega_c\right]^{-1} {\diff \omega \over \omega} \quad \hbox{with} \quad 
\omega \geq \omega_c, \eeq
where $v_c$ and $\omega_c$ (and $k_c$ below) characterize the largest 
convective eddies. 
We invoke once more Kolmogorov's law to replace $k^2(\omega)$ in 
(\ref{distrik}) by $k_c^2(\omega / \omega_c)^3$, and reach the following
expression for the flux of kinetic energy at the top of the radiative
interior:  
\beq 
{\cal F}_K(r_c) = \rho_c v_c^3
{\omega_c \over N_c^2}
\int_{\omega_c}^{N_c} \! {\diff \omega \over \omega}\; 
(N_c^2-\omega^2)^{1\over2} Ê\left[{\omega \over {\omega_c}}\right]^{-2} \!
\int_0^{\ell} {\diff \ell \over \ell_c}. 
\eeqn{kespect}
From now on we use the spherical harmonic number $\ell = r k_h$,
and pretend that it varies continuously from $0$ to its upper bound
$\ell_u$ (the convective scale):
\beq
0< \ell < \ell_u \quad \hbox{with} \; \ell_u = \ell_c \left({\omega \over
{\omega_c}}\right)^{3\over2}. 
\eeqn{limell} 

In the limit ${\omega_c} \ll N_c$ the flux integrated over the
whole spectrum amounts to 
\beq 
{\cal F}_K(r_c) = 2 \, \rho v_c^3 \left({{\omega_c} \over N_c}\right) ,
\eeq
a result which is similar to that given by Press for his monochromatic flux [cf.
his eq. (90)]. Note however that our expression
(\ref{kespect}) for  ${\cal F}_K$ vanishes when ${\omega_c} \rightarrow N_c$,
whereas his is singular there.

To proceed with the evaluation of the angular momentum flux, we need to 
know how the spectral energy of the internal waves is distributed over  
the azimuthal order $m$, for given $\ell$. For simplicity, we shall assume
that this distribution is uniform, although it is quite possible that the
Coriolis force, which we have neglected here, causes an unbalance
bet\-ween prograde and retrograde waves, hence between positive and
negative $m$. With our simplifying assumption  
\beqan 
{\cal F}_K(r_c) &=& \rho_c v_c^3 {{\omega_c} \over N_c^2} \\
&\times&\int_{\omega_c}^{N_c} \! {\diff \omega \over \omega}\; 
(N_c^2-\omega^2)^{1\over2} Ê\left[{\omega \over {\omega_c}}\right]^{-2} \!
\int_0^{\ell_u} {\diff \ell \over \ell_c} \int_{-\ell}^{\ell}          
{\diff m \over 2 \ell}. \nonumber 
\eeqan{kespectm}
Referring back to (\ref{fluxj}) and (\ref{dampj}), we obtain the following 
expression for the luminosity of angular momentum 
integrated over the whole wave spectrum: 
\beqan 
\lefteqn{{\cal L}_J(r) =  4 \pi r^2 \frac{\rho_c v_c^3}{N_c \ell_c}}
\label{jspectm} \\
\lefteqn{\times \! \int_{\omega_c}^{N_c} \! {\diff \omega \over \omega}\! 
\left(1-{\omega^2 \over N_c^2}\right)^{1\over2} \!\! \left[{\omega \over
{\omega_c}}\right]^{-3}  \!\! \int_0^{\ell_u} {\diff \ell \over \ell} 
\int_{-\ell}^{\ell} \! \exp(-\tau) \;  m \, \diff m . }
\nonumber
\eeqan{nono}
In a stationary regime, one would have to subtract from this luminosity that, 
of opposite sign, which is carried by the waves reflected at the center, 
including the appropriate damping. 

\section{The angular momentum flux}
In a non-rotating star, there is no net flux of angular momentum, since
two waves propagating in the same direction but which are of opposite
azimuthal order $m$ have the same amplitude;  therefore they carry (and 
deposit) equal amounts of
angular momentum, which cancel each other because they are of opposite sign. 

The situation is different in a rotating star, because
the Coriolis force may introduce a bias between waves of opposite $m$, and
this could generate differential rotation as the waves undergo radiative 
damping.  We do not consider this possibility here, and assume rather that
differential rotation exists for another reason, such as mass loss or
meridian circulation. In this case, waves of opposite azimuthal order 
experience different damping, and the net effect will be an extraction of 
angular momentum from the radiation zone.
When the damping is slight, the associated flux of angular momentum
will be small, and we shall neglect it in this first approach. 
Instead, we concentrate on those waves which are completely damped
because their local frequency vanishes at some critical depth. A similar 
damping process plays an important role in the tidal synchronization of
massive binaries, as was pointed out by Goldreich \& Nicholson (1989).

The local frequency may be written   
\beq 
\sigma(r) = \omega - m \, [\Omega(r) -
{\Omega_c}] \equiv \omega - m \; \Delta \Omega(r) 
\eeqn{doppler2} 
with $\omega$ being
the frequency of the monochromatic wave when it is emitted at the base of 
the convection zone, which rotates at the angular velocity ${\Omega_c}$.
If the rotation speed $\Omega(r)$ increases
with depth, $\sigma$ will have a node at some
location $r^*(\omega, m)$ for sufficiently large (and positive) $m$.
There $k_v \rightarrow \infty$ and radiative damping will increase 
dramatically --  formally $\tau \rightarrow \infty$, but the
quasi-adiabatic approximation leading to (\ref{integral}) is no longer
valid. Thus at the critical layer $r=r^*$ this wave will deposit whatever 
remains from its (negative) angular momentum. 
The result is similar if the rotation rate
decreases with depth, but it is then of opposite sign; in each case radiative
dissipation acts to reduce the existing differential rotation.

From now on, we shall consider only those waves which experience little
damping before they reach the critical level $r^*$ where $\sigma \rightarrow
0$;   according to (\ref{integral}), their initial frequency $\omega$ is such 
that $\tau \la 1$, or 
\beq
\omega^4 \ga I(r^*)\, \ell^3 \quad \hbox{where} \;\;
I(r) = \int_r^{r_c} K  \; N \, N_t^2 \, {\diff r \over r^3}.
\eeqn{integral2}
A wave which does not meet this condition is damped before, but its
contribution to the transport of angular momentum is  compensated by    
that of its partner of opposite $m$ and we shall ignore it in this first
approach.
\begin{figure}[t]
\centerline{
\epsfxsize=8.7cm
\epsfysize=9.0cm
\epsfbox{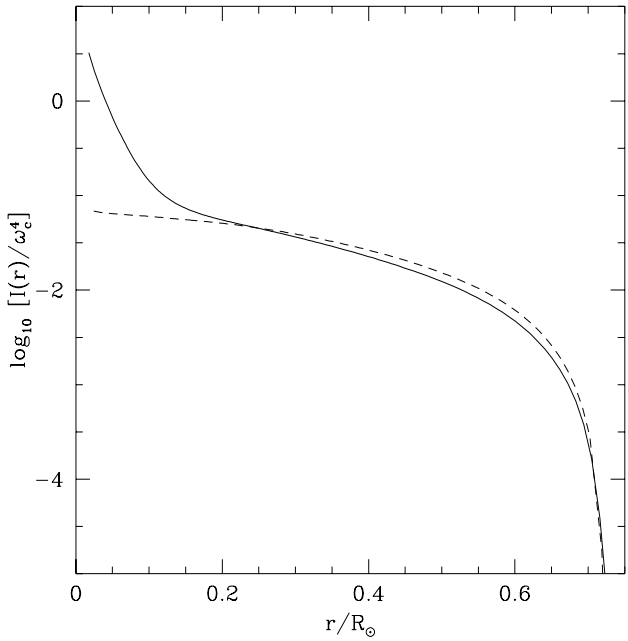}
}
\caption{Variation with depth of the damping integral defined in
(31), scaled by the fourth power of the convective
turnover frequency: $I/\omega_c^4$. The dashed line refers to the Sun
at 200 Myr, the continuous line to the present Sun. The radial cooordinate is  normalized by the actual, time-dependent radius of the Sun. }  
\end{figure}

A final, admittedly drastic simplification will enable us to calculate 
the triple integral (\ref{jspectm}): we shall approximate the damping
exponential by the step function ${\cal H}(r-r^*)$.  We thus take
$\exp[-\tau(r)] = 1$ for $\omega > m \, \Delta \Omega$ and     
$\exp[-\tau(r)] = 0$ for $\omega < m \, \Delta \Omega$. In other words,
$\exp[-\tau]$ will be removed from (\ref{jspectm}), and the integration
domain in $m$ will be delimited by  
\beq
0 < m < {\rm min} [\ell, {\omega / \Delta \Omega}] .
\eeq
Let us recall the other limits; according to
(\ref{integral2}) and  (\ref{limell})
\beq
0 < \ell < {\rm min} [(\omega^4 /I)^{1\over3}, \ell_c 
(\omega/{\omega_c})^{3\over2}] 
\eeqn{codil}
and for the frequency we have
\beq
{\omega_c} < \omega < N_c .
\eeqn{codiom}

The integration can now be performed, and its results are reported in Appendix A.
The angular momentum flux is an  implicit function of $r$, through
$\Delta \Omega(r)$ and  $I(r)$, whose variation with depth is displayed
in Fig.~1. Its general expression is rather intricate, but it simplifies
in some parameter domains. For instance, let us take the interval in $I$ 
which is the most relevant for the solar interior, namely where 
\beq 
\ell_c^{-3} < \left({I \over {\omega_c}^4}\right) < 1 \,
\eeq
and that in $\Delta \Omega(r)$ defined by 
\beq
{{\omega_c} \over N_c} \left({I \over {\omega_c}^4}\right) < 
\left({\Delta \Omega \over {\omega_c}}\right)^3 <
\left({I \over {\omega_c}^4}\right)  
\eeq
(for the physical meaning of these limits we refer the reader to Appendix A.)
In this domain the flux depends linearly on the differential rotation:
\beq
{\cal F}_J = {\rho_c v_c^3 \over N_c \ell_c} 
\left[{3 \over 4} \left({{\omega_c}^4 \over I}\right)^{2 \over 3}
- {1 \over 3} \left({{\omega_c}^4 \over I}\right) 
{\Delta \Omega \over {\omega_c}}\right] .
\eeqn{resjflux}
The term independent of $\Delta \Omega$ represents the contribution of 
waves which do not meet the singularity at $\sigma = 0$, and consistent with
the approximation made above, we shall ignore it. We thus write the result
as  
\beq
{\cal L}_J(r) = {\cal L}_J(r_c) - {4 \pi r^2 \over 3}                
{\rho_c v_c^3 \over N_c \ell_c}  \left({{\omega_c}^4 \over I}\right)   
{\Delta \Omega \over {\omega_c}}. 
\eeqn{corjflux}

\section{Efficiency of the angular momentum transport}
If angular momentum is transported only by the internal waves we have
considered here, the angular velocity evolves with time according  to 
\beq
{\partial \over \partial t} \int\!\!\!\int \Omega r^2  \sin^2 \theta\;
\rho r^2 \sin \theta  \, \diff \theta \, \diff \phi = - {\partial \over
\partial r} {\cal L}_J(r)   \eeq
or equivalently
\beq
{\partial \over \partial t} \bigl( \rho r^4 \Omega \bigr) =
{1 \over 2} {\rho_c v_c^3 \over N_c \ell_c} \; {\partial \over \partial r}
\left[r^2 \left({{\omega_c}^4 \over I}\right)  {\Delta \Omega \over {\omega_c}}
\right]. 
\eeq

To evaluate the efficiency of this transport, we neglect all 
variations except those of $\Omega$; hence
\beq
{\partial \Omega \over \partial t} \approx  V_{\rm w} 
{\partial \Omega \over \partial r} ,
\eeqn{front}
whose formal solution is
\beq
\Omega(r,t) = F \left( r + \int V_{\rm w} \, \diff t \right) .
\eeq
Within this approximation the rotation profile, whose slope 
depends on the rate of angular momentum loss through the wind,
propagates inwards with the velocity $V_{\rm w}$, which is directly
related to the damping integral $I(r)$ defined in (\ref{integral2}): 
\beq
V_{\rm w} =  {1 \over 2} {\rho_c v_c^3 \over \rho r^2}{1 \over N_c
{\omega_c} \ell_c} \left({{\omega_c}^4 \over I}\right).
\eeq 

Approximating $\rho_c v_c^3$ by 1/10 of the convective flux  (see Cox \&
Giuli 1968), and this flux by $L_\odot/4 \pi r_c^2$, we obtain a
crude estimate for the synchronization time $t_{\rm sync} = 
{r / V_{\rm w}}$: 
\beq
t_{\rm sync} 
\approx 60 \,{M_\odot R_\odot^2 \over L_\odot}  \,   {\rho \over \overline
\rho} \left({r \over R_\odot}\right)^3 \! \left({r_c \over
R_\odot}\right)^2 \! N_c {\omega_c} \ell_c \left({I \over
{\omega_c}^4}\right),   \eeq
$\overline \rho$  being the mean density.
The variation with depth of $t_{\rm sync}$ 
is depicted in Fig.~2; for this evaluation we took $\omega_c=2 \pi  v_c/H_P$, $\ell_c=2
\pi r_c/H_P$, and assumed a penetration depth of $H_P/10$, which
determines the value of $N_c$ ($\approx 10^{-3}$ s$^{-1}$). The solar
models were built with the stellar evolution code CESAM (Morel 1996).  
\begin{figure}[t]
\centerline{
\epsfxsize=8.7cm
\epsfysize=9.0cm
\epsfbox{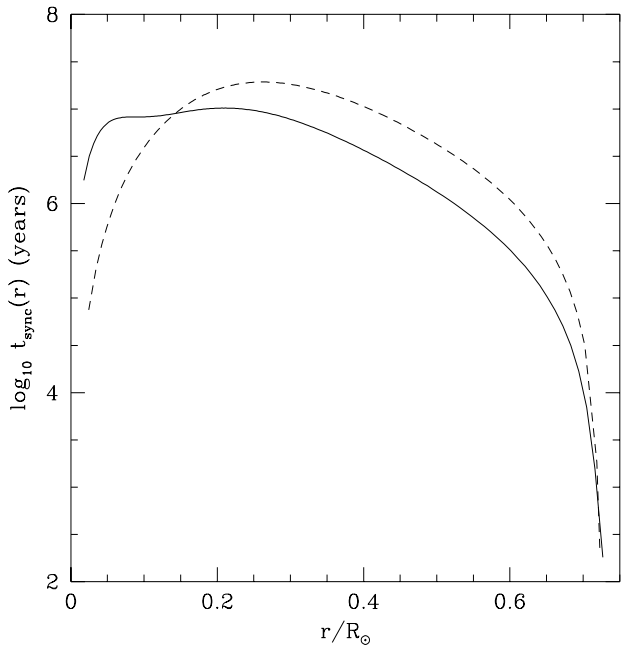}
}
\caption{Variation with depth of the synchronization time 
$t_{\rm sync}=  r/V_w$ (see eq. 44). As in Fig.~1, the dashed line refers to
the Sun at 200 Myr, the continuous line to the present Sun.}
\end{figure}

The synchronization time depends somewhat on the age of the Sun,
because the damping of the internal waves increases with time in the
inner core, due to the steepening of the molecular weight gradient, but
this effect is compensated by the increase of the convective frequency
$\omega_c$.  We see that the internal waves are able to extract angular
momentum from the solar interior in about $10^7$ years, three orders of
magnitude less than the present spin-down time.

\section{Conclusion}
We conclude that the internal gravity waves are quite efficient in
extracting angular momentum from the radiative interior of the Sun, a
result anticipated by Schatzman (1993).
During the  later stages of the solar spin-down, this
process prevails over the other mechanisms which have been proposed so
far, except perhaps magnetic torquing  (see Introduction). 
The efficiency of this transport has been 
confirmed independently by Kumar \&
Quataert (1996),
but these authors considered the differential damping between waves
of opposite azimuthal order $m$, which redistributes angular momentum
within the radiation zone with no net flux into the convective envelope. 
In contrast, we neglected here this effect altogether and focussed
only on those waves which undergo complete damping at
the critical level where their
local frequency vanishes.
We are
implementing this transport in our rotational evolution codes, where it
will compete with other mechanisms, such as meridian circulation, and we
shall report the results in forthcoming papers.

But we clearly need to move beyond the present exploration, whose weak
points ought to be recalled. First we have excluded the Coriolis force
from the description of the internal waves, thus missing a possible
unbalance between waves of opposite azimuthal order $m$. 
Second, we neglected the redistribution of angular momentum due to
the differential damping mentioned above.
Finally we used a rather crude recipe to couple the internal waves with
the convective motions, which we assumed to obey Kolmogorov's law; it
could well be that the stochastic excitation at wavelengths exceeding the
size of the convective eddies is overestimated, but the result is not too
sensitive to the slope of the power spectrum. 

In spite of these imperfections, which have to be remedied, we see only one
interpretation of the short characteristic time we have found above. 
Namely that the transport of angular momentum through the internal waves
which are emitted by the convection zone is responsible for the flat rotation
profile which is observed in the Sun. 
\goodbreak

\begin{acknowledgements}
Shortly before submitting this paper, we learned that P. Kumar was
working on the same problem. The exchanges which followed, and several
discussions with E. Schatzman, incited us to clarify some delicate points.   
We thank P. Morel for allowing us to use his stellar evolution code
CESAM. J.M. was supported by a grant Ciencia/Praxis from JNICT of
Portugal, and S.T. gratefully acknowledges support from NSERC of Canada.  
\end{acknowledgements} 

\bigskip

\noindent {\bf Appendix A: Expressions for the angular momentum flux}

\bigskip
\noindent The angular momentum luminosity is given by eq. (\ref{jspectm}):
\beqm
{\cal L}_J(r) =  4 \pi r^2 {\rho_c v_c^3 \over N_c \ell_c}  \; Q(r)
\nonumber 
\eeqm
with
\beqm
Q(r) = \int_{\omega_c}^{N_c} \! {\diff \omega \over \omega}\!
\left(1-{\omega^2 \over N_c^2}\right)^{1\over2} \!\! Ê\left[{\omega \over
{\omega_c}}\right]^{-3}  \!\! \int_0^{\ell_u} {\diff \ell \over \ell} 
\int_0^{\ell} \!   m \, \diff m .
\eeqm
Let us recall that we have replaced the summations over integer $m$ and $\ell$ by
integrals over these quantities, which are assumed to vary continuously.
We take $m=1$ (and therefore $\ell=1$) as lowest value when defining
validity domains for various expressions. But we shall keep $0$ as lower bound
for the integrals, because for small values of $L$ they approximate
better sums such that 
\beqm  
\sum_1^L {1 \over \ell} \, \sum_1^\ell m = {1\over4} L^2. 
\eeqm 
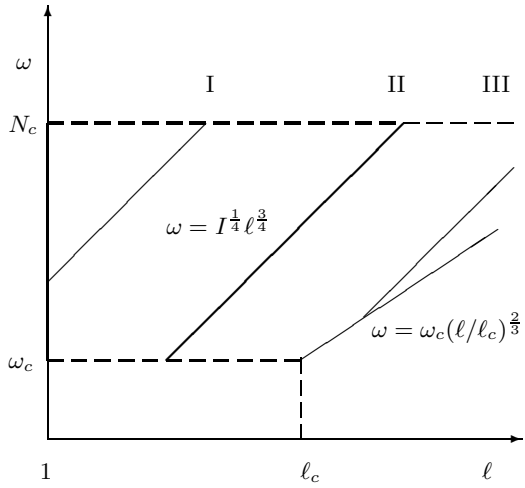
\begin{figure}[t]
        \setlength\unitlength{1.05cm}
\centerline{
\begin{picture}(7.7,6.5)
        \put(1,1){\vector(0,1){5.5}}
        \put(1,1){\vector(1,0){6.0}}
        \multiput(1,2)(0.3,0){11}{\line(1,0){0.19}}
        \multiput(4.2,1)(0,0.28){4}{\line(0,1){0.19}}
        \multiput(1,5.01)(0.3,0){20}{\line(1,0){0.2}}
        \put(4.2,2){\line(3,2){2.5}}
        \put(1,3){\line(1,1){2.0}}
        \put(5,2.55){\line(1,1){1.9}}
        \thicklines
        \multiput(1,2)(0.3,0){5}{\line(1,0){0.19}}
        \multiput(1,5.01)(0.3,0){15}{\line(1,0){0.2}}
        \put(1,2){\line(0,1){3.0}}
        \put(2.50,2){\line(1,1){3.0}}
        \thinlines
        \put(0.9,0.5){$1$}
        \put(4.2,0.5){$\ell _c$}
        \put(6.5,0.5){$\ell$}
        \put(0.5,1.9){$\omega _c$}
        \put(0.5,4.9){$N_c$}
        \put(0.6,5.7){$\omega $}
        \put(2.5,3.6){$\omega = I^{1\over4} \ell^{3\over4}$}
        \put(5.1,2.3){$\omega =\omega_c (\ell/\ell_c)^{2\over3}$}
        \put(3.0,5.4){I}
        \put(5.3,5.4){II}
        \put(6.5,5.4){III}
\end{picture}
}
\caption{Limits of the integration domain in the $(\ell, \omega)$ plane, in
logarithmic coordinates. In case II, which is the most relevant to the solar
interior, the damping line $\omega = I^{1/4} \ell^{3/4}$ cuts $\omega =
\omega_c$ between $1$ and $\ell_c$; the corresponding integration domain is 
delineated in thick lines.}  
\end{figure} 
We calculate first the angular momentum flux in the range  
\beqm
\ell_c^{-3} < \left({I \over \omega_c^4}\right) < 1
\eeqm
which corresponds to case II in Fig.~3; it covers most of the existence
domain of the integral $I(r)$, as shown in Fig.~1. Then the
upper limit of $\ell$, for given $\omega$, is fixed by condition
(\ref{codil})  \beqm
\ell_u = \left({\omega^4 \over I}\right)^{1\over 3}
\eeqm
and the integration domain is split in two, according to whether
$\ell_u$ is smaller or larger than $\omega /\Delta \Omega$ (see Fig.~4),
or equivalently whether $\omega$ is smaller or larger than $\omega'$, 
\beqm
\hbox{with} \qquad {\omega' \over \omega_c} = {I \over\omega_c^4}
\left({\omega_c \over \Delta \Omega}\right)^3 .
\eeqm
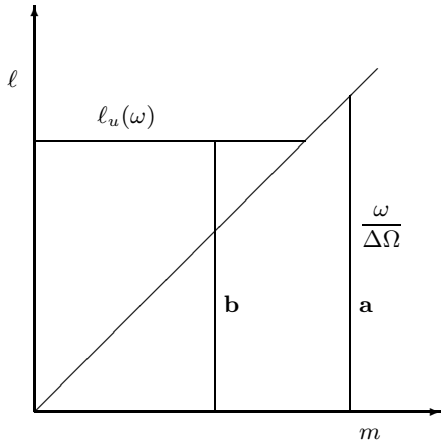
\begin{figure}[t]
        \setlength\unitlength{1.2cm}
\centerline{
\begin{picture}(6,6)
        \put(1,1){\vector(0,1){4.5}}
        \put(1,1){\vector(1,0){4.5}}
        \put(1,1){\line(1,1){3.8}}
        \put(1,4){\line(1,0){3}}
        \put(3,1){\line(0,1){3}}
        \put(4.5,1){\line(0,1){3.5}}
        \put(4.6,0.7){$m$}
        \put(0.7,4.6){$\ell$}
        \put(1.7,4.2){$\ell_u (\omega)$}
        \put(3.1,2.1){\bf b}
        \put(4.6,2.1){\bf a}
	\put(4.6,3){$\displaystyle{\omega \over \Delta \Omega}$}
\end{picture}
}
\caption{Limits of the integration domain in the $(m, \ell)$ plane.}
\end{figure}
\goodbreak
\medskip
\noindent {\it Subdomain} {\bf a} : $\; \omega_c < \omega < \omega'$
(or $\ell_u < \omega / \Delta \Omega$) 
\medskip

In the $(m, \ell)$ plane the integration domain is a triangle:
\beqm
P_a(\omega) = \int_0^{\ell_u} {\diff \ell \over \ell}
\int_0^\ell m \,\diff m = 
{1 \over 4} \left({\omega_c^4 \over I}\right)^{2\over 3}
\left({\omega \over \omega_c}\right)^{8\over3}
\eeqm
and therefore
\beqan
Q_a &=& \int_{\omega_c}^{\omega'} P_a(\omega)
\left({\omega \over \omega_c}\right)^{-3}
{\diff \omega \over \omega} \nonumber \\
&=&{1 \over 4 } \left({\omega_c^4 \over I}\right)^{2\over3}
\int_{\omega_c}^{\omega'} \left({\omega \over \omega_c}\right)^{-{1\over3}}
{\diff \omega \over \omega} \nonumber \\
&=& {3 \over 4} \left({\omega_c^4 \over I}\right)^{2\over3}
\left[1 - {\Delta \Omega \over \omega_c} 
\left({\omega_c^4 \over I}\right)^{1\over3} \right] .
\eeqan{qa}
\goodbreak
\medskip
\noindent {\it Subdomain} {\bf b} : $\; \omega' < \omega < N_c$
(or $\ell_u > \omega / \Delta \Omega$) 
\medskip

The integration domain in $(m, \ell)$ is now a trapezium; for the
triangular tip we have
\beqm
P_{b1}(\omega) = \int_0^{\omega / \Delta \Omega} {\diff \ell \over \ell}
\int_0^\ell m \,\diff m = 
{1 \over 4} \left({\omega \over \Delta \Omega}\right)^2
\eeqm
and in the rectangular part
\beqm
P_{b2}(\omega) = \!\int_{\omega / \Delta \Omega}^{\ell_u} {\diff \ell \over
\ell} \int_0^{\omega / \Delta \Omega}\!\!\!\! m \,\diff m = 
{1 \over 2} \left({\omega \over \Delta \Omega}\right)^2
\!\ln \left({\ell_u \over \omega / \Delta \Omega}\right)\!.
\eeqm
Summing the two we get
\beqa
P_{b}=P_{b1}+P_{b2}&=& {1 \over 4}  
\left({\omega_c \over \Delta \Omega}\right)^2
\left({\omega \over \omega_c}\right)^2 \nonumber \\
& \times& \left[1 - {2 \over3} \ln {I \over \omega_c^4} 
+2 \ln {\Delta \Omega \over \omega_c }
+ {2 \over3} \ln {\omega \over \omega_c} \right]. \nonumber
\eeqa

It remains to integrate over $\omega$; taking into account that
$\omega_c \ll N_c$, we have
\beq
Q_b = \int_{\omega'}^{N_c} P_b(\omega)
\left({\omega \over \omega_c}\right)^{-3}
{\diff \omega \over \omega}
= {5 \over 12} {\Delta \Omega \over \omega_c} 
\left({\omega_c^4 \over I}\right).
\eeqn{qb}
and hence
\beq
Q(r)=Q_a + Q_b = {3 \over 4} \left({\omega_c^4 \over I}\right)^{2\over3}
-{1 \over 3} \left({\omega_c^4 \over I}\right) 
{\Delta \Omega \over \omega_c}  , \label{RII.2}
\eeq
which is the result quoted above in eq. (\ref{resjflux}).
This expression is valid provided
$\omega_c < \omega' < N_c$,  in other terms as long as
\beq
{{\omega_c} \over N_c} \left({I \over {\omega_c}^4}\right) < 
\left({\Delta \Omega \over {\omega_c}}\right)^3 <
\left({I \over{\omega_c}^4}\right). 
\eeq

So far, we have implicitly assumed that
$ \omega_c <\omega' < N_c$; let us examine the other cases.

For $\omega' > N_c$, or
\beq
\left({\Delta \Omega \over {\omega_c}}\right)^3 <
{{\omega_c} \over N_c} \left({I \over {\omega_c}^4}\right)
\eeq
$Q_b=0$ and $Q(r)$ does not depend on $\Delta \Omega$ anymore
\beq
Q(r)=Q_a = {3 \over 4} \left({\omega_c^4 \over I}\right)^{2\over3}
\left[1- \left({\omega_c \over N_c}\right)^{1\over3}\right] . \label{RII.1}
\eeq

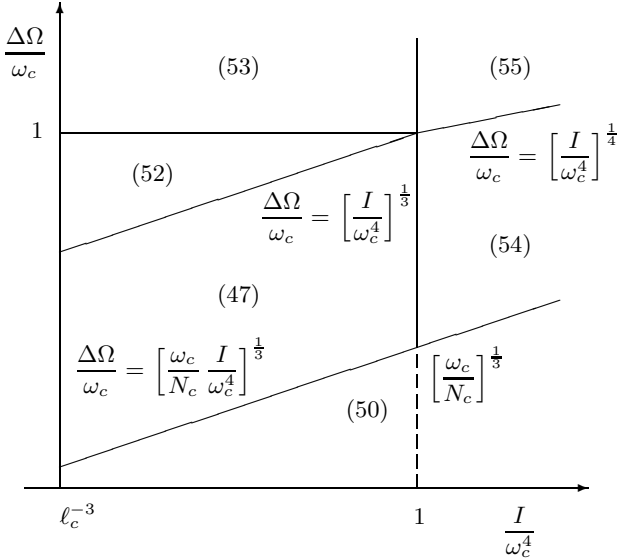
\begin{figure}[t]
        \setlength\unitlength{0.95cm}
\centerline{
\begin{picture}(9.0,8)
        \put(6,3.2){\line(0,1){4.3}} 
        \multiput(6,1.2)(0,0.28){8}{\line(0,1){0.19}}
        \put(0.5,1.2){\vector(1,0){7.9}} 
        \put(1,1.2){\vector(0,1){6.8}} 
        \put(1,1.5){\line(3,1){7.0}} 
        \put(1,4.5){\line(3,1){5.0}}
        \put(6,6.17){\line(5,1){2.0}}
        \put(1,6.17){\line(1,0){5.0}}
        \put(1.0,0.7){$\ell_c^{-3}$}
        \put(5.95,0.7){$1$}
        \put(6.15,2.65){$\displaystyle{\left[{\omega_c 					        \over N_c}\right]^{1\over3}}$}
        \put(1.2,2.75){$\displaystyle{\Delta \Omega \over \omega_c}=$}
        \put(2.3,2.75) {$\displaystyle{\left[{\omega_c \over N_c }				\,{I \over \omega_c^4}\right]^{1\over3}}$}
        \put(3.8,4.9){$\displaystyle{\Delta \Omega \over \omega_c}=$}
        \put(4.9,4.9) {$\displaystyle{\left[{I \over \omega_c^4}\right]
		     ^{1\over3}}$}
        \put(6.7,5.75){$\displaystyle{\Delta \Omega \over \omega_c}=$}
        \put(7.8,5.75) {$\displaystyle{\left[{I \over \omega_c^4}\right]
		     ^{1\over4}}$}
        \put(0.2,7.2){$\displaystyle{\Delta \Omega \over \omega_c}$}
        \put(0.6,6.1){$1$}
        \put(7.2,0.5){$\displaystyle{I \over \omega_c^4}$}
        \put(5.0,2.2){(\ref{RII.1})}
        \put(3.2,3.8){(\ref{RII.2})}
        \put(2.0,5.5){(\ref{RII.3})}
        \put(7.0,4.5){(\ref{RI.2})}
        \put(7.0,7.0){(\ref{RI.3})}
        \put(3.2,7.0){(\ref{RII.4})}
\end{picture}
}
\caption{The regions in the $(I/\omega_c^4, \Delta \Omega /\omega_c)$ plane
where the angular momentum flux is given respectively by expressions 
(\ref{RII.1}), (\ref{RII.2}), (\ref{RII.3}), (\ref{RII.4}),
(\ref{RI.2}) and (\ref{RI.3}).
Logarithmic coordinates have been used again,
but scales are arbitrary.}  
\end{figure}

Last we turn to the case where $\omega' < \omega_c$, or
\beq 
\left({I \over {\omega_c}^4}\right) <
\left({\Delta \Omega \over {\omega_c}}\right)^3 . 
\eeq
Then $Q_a = 0$, and
the integration of $P_b$ must be performed over the domain
min$(\omega_c, \Delta \Omega) < \omega < N_c$.
The result is (again with $\omega_c \ll N_c$)
\beq
Q(r)=\left({\omega_c \over \Delta \Omega}\right)^2
\left[{5 \over 12} - {1 \over 6}\ln {I \over \omega_c^4}
+ {1 \over 2}\ln {\Delta \Omega \over \omega_c} \right] \label{RII.3}
\eeq
for $\Delta \Omega < \omega_c$, and
\beq
Q(r)=\left({\omega_c \over \Delta \Omega}\right)^3
\left[{5 \over 12} - {1 \over 6}\ln {I \over \omega_c^4}
+ {2 \over 3}\ln {\Delta \Omega \over \omega_c} \right] \label{RII.4}
\eeq
for $\Delta \Omega > \omega_c$. 
\medskip

Finally we have to consider the case I of Fig.~3, where $I>\omega_c^4$,
which applies to the innermost core of the evolved Sun (see Fig.~1).
Performing the integrations as above, one finds 
\beq
Q(r)=Q_a + Q_b = {3 \over 4} \left({\omega_c^4 \over I}\right)^{3\over4}
-{1 \over 3} \left({\omega_c^4 \over I}\right) 
{\Delta \Omega \over \omega_c}  , \label{RI.2}
\eeq
for $(\Delta \Omega / \omega_c)^4 < I/\omega_c^4$, and
\beq
Q(r)= \left({\omega_c^4 \over I}\right)^{1\over4}
\left({\omega_c \over \Delta \Omega}\right)^2 \left[{5 \over 12} - 
{1 \over 8}\ln{I \over \omega_c^4} + {1 \over 2}\ln {\Delta \Omega \over
\omega_c} \right]  \label{RI.3}
\eeq
when $(\Delta \Omega / \omega_c)^4 > I/\omega_c^4$.

It is easy to check that  $Q(r)$
given successively by (\ref{RII.1}), (\ref{RII.2}), (\ref{RII.3}), (\ref{RII.4}),
(\ref{RI.2}) and (\ref{RI.3})
is a continuous function of $\Delta \Omega (r)$ and $I(r)$, and hence of $r$.
In Fig.~5, we display the domains where each of these expressions applies.
\goodbreak

\bigskip
\noindent {\bf Appendix B: Effect of a molecular weight gradient}

\bigskip
\noindent A molecular weight gradient modifies somewhat the dynamics of
internal waves, and in particular their radiative damping. To take this into
account, we follow the method adopted by Press (1981), and refer to his
equations as  (P~.~.~).

The equation of motion is still (P 27) 
\beq
i \sigma \nabla^2 (\rho u_r) = g k_h^2 \rho'
\eeqn{motion1}
but the equation of state now includes the Eulerian perturbation of the
molecular weight $\mu$~:
\beq
{\rho' \over \rho} = - \delta {T' \over T} + \varphi {\mu' \over \mu},
\eeqn{etat}
where we have ignored the pressure perturbation, as allowed for internal
waves (anelastic approximation). The fluctuation of $\mu$ is given by
\beq
 i \sigma {\mu'\over \mu} = {\nabla_\mu \over H_P}\, u_r ,
\eeqn{consmu}
which expresses the conservation of molecular weight in the wave motion. 

To eliminate $T'$, we call the heat equation
\beq
i \sigma {T' \over T} = - (\nabla_{\rm ad} - \nabla) \, {u_r\over H_P} +
{K \over T} \nabla^2 T' ,
\eeq
$K$ being the thermal diffusivity;
using (\ref{etat}) and (\ref{consmu}), and neglecting the vertical variation
of all mean quantities compared to that of $T'$, we get
\beq
i \sigma {\rho' \over \rho} = {N^2 \over g} u_r - 
K \nabla^2 \left( \delta {T' \over T} \right) , 
\eeq
where
\beqm
N^2 = N_t^2 + N_\mu^2 = 
{g \delta \over H_P} (\nabla_{\rm ad} - \nabla) + 
{g  \varphi \over H_P} \nabla_\mu  .\nonumber 
\eeqm

Next we draw $(T' / T)$ from the equation of state (\ref{etat}) to
obtain 
\beq
(i \sigma - K \nabla^2) \rho' = {N^2\over g} \rho u_r 
- {K \over i \sigma} \nabla^2 \left( {N_\mu^2 \over g} \rho u_r \right)
\eeqn{rhop}
which is (P 23) plus the extra term in $N_\mu^2$.
Combining (\ref{motion1}) and (\ref{rhop}), we are led to
\beq
\left( \nabla^2 + k_h^2 {N^2 \over \sigma^2} \right) \rho u_r
+ i {K \over \sigma}  \nabla^2
\left( \nabla^2 + k_h^2 {N_\mu^2 \over \sigma^2} \right) \rho u_r 
= 0 ,
\eeq
which again differs from (P 28) by this term in $N_\mu^2$.
\goodbreak

This equation is readily transformed into the
dispersion relation
\goodbreak
\beqa
\biggl[ k_v^2 &-&   k_h^2 \left({N^2 \over \sigma^2} - 1 \right)
\biggr]  \\ &-&  i \, {K \over \sigma} \left[k_v^2 + k_h^2 \right]
\left[k_v^2 - k_h^2 \left({N_\mu^2 \over \sigma^2} - 1 \right)\right] = 0 ,
\nonumber  
\eeqa
which yields the following expression for the vertical wavenumber, in
the quasi-adiabatic limit:
\beq 
k_v = \pm k_h \! \left({N^2 \over \sigma^2} - 1 \right)^{1 \over 2} \!  +   
i \, {K \over 2\sigma} \, k_h^3 \, { N \, N_t^2  \over \sigma^3} 
\left({N^2 \over N^2 - \sigma^2} \right)^{1 \over 2} , 
\eeq
and hence the value of the damping integral $\tau(r)$ quoted above in
(\ref{integral}):
\beqm
\tau(r) = [\ell(\ell+1)]^{3/2} \int_r^{r_c} K  \; {N \, N_t^2 \over
\sigma^4}  \left({N^2 \over N^2 - \sigma^2}\right)^{1 \over 2} {\diff r
\over r^3}.  \nonumber
\eeqm


\bigskip


\end{document}